\documentclass[10pt,conference]{IEEEtran}

\newif\ifSPACEHACK
\SPACEHACKtrue 

\newif\ifDEBUG
\DEBUGtrue

\newif\ifANONYMOUS
\ANONYMOUSfalse

\usepackage{amsmath,amssymb,amsfonts}
\usepackage{algorithmic}
\usepackage{graphicx}
\usepackage{textcomp}
\usepackage{xcolor}
\usepackage{booktabs} 
\usepackage{xspace} 
\usepackage[normalem]{ulem}
\usepackage{makecell}
\usepackage{tcolorbox}
\usepackage{enumitem}
\usepackage{siunitx}
\usepackage[vskip=1em,font=itshape,leftmargin=2em,rightmargin=2em]{quoting}
\usepackage{hyperref}

\usepackage[noadjust]{cite}

\usepackage{soul}
\usepackage{multirow}
\usepackage{array,booktabs}
\newcolumntype{M}[1]{>{\centering\arraybackslash}m{#1}}
\usepackage{amsfonts}
\usepackage{amsmath}
\usepackage{wrapfig, lipsum}
\usepackage{subfigure}

\ifDEBUG
    \newcommand{\JD}[1]{\textcolor{purple}{[JD:#1]}}
    \newcommand{\AG}[1]{\textcolor{olive}{[AG:#1]}}
    \newcommand{\WJ}[1]{\textcolor{blue}{[WJ:#1]}}
    \newcommand{\GKT}[1]{\textcolor{brown}{[GKT:#1]}}
    \newcommand{\MATT}[1]{\textcolor{pink}{[MH:#1]}}
    \newcommand{\NMS}[1]{\textcolor{orange}{[NMS:#1]}}
    \newcommand{\WPM}[1]{\textcolor{red}{[Trey: #1]}}
    \newcommand{\TRS}[1]{\textcolor{teal}{[Taylor: #1]}}
    \newcommand{\AT}[1]{\textcolor{magenta}{[AT: #1]}}
    \newcommand{\CS}[1]{\textcolor{purple}{[CS: #1]}}
    
    \newcommand{\TODO}[1]{\hl{#1}}
\else
    \newcommand{\JD}[1]{}
    \newcommand{\AG}[1]{}
    \newcommand{\WJ}[1]{}
    \newcommand{\GKT}[1]{}
    \newcommand{\NS}[1]{}
    \newcommand{\NV}[1]{}
    \newcommand{\NMS}[1]{}
    \newcommand{\WPM}[1]{}
    \newcommand{\TRS}[1]{}
    \newcommand{\MATT}[1]{}
    \newcommand{\AT}[1]{}
    
    \newcommand{\TODO}[1]{}
\fi

\newif\ifSPACEHACK
\SPACEHACKfalse

\usepackage[compact]{titlesec}
    \usepackage{etoolbox}
    \makeatletter
    \patchcmd{\ttlh@hang}{\parindent\z@}{\parindent\z@\leavevmode}{}{}
    \patchcmd{\ttlh@hang}{\noindent}{}{}{}
    \makeatother

\ifSPACEHACK
    \titlespacing*\section{0pt}{5pt plus 1pt minus 1pt}{3pt plus 1pt minus 1pt}
    \titlespacing*\subsection{0pt}{4pt plus 1.5pt minus 1.5pt}{4pt plus 1.5pt minus 1.5pt}
    \titlespacing*\subsubsection{0pt}{3pt plus 1pt minus 1pt}{3pt plus 1.5pt minus 1.5pt}
    \titlespacing*\paragraph{0pt}{1pt plus 1.5pt minus 1.5pt}{2pt plus 1.5pt minus 1.5pt}
\fi
\usepackage{balance} 

\usepackage{amsmath}
\usepackage{cleveref}

\crefformat{section}{\S#2#1#3}
\crefname{figure}{Figure}{Figures}
\crefname{appendix}{Appendix}{Appendices}
\crefname{table}{Table}{Tables}
\crefname{algorithm}{Algorithm}{Algorithms}
\crefname{listing}{Listing}{Listings}
\crefname{theorem}{Theorem}{Theorems}
\crefname{thm}{Theorem}{Theorems}
\crefname{lemma}{Lemma}{Lemmata}
\crefname{equation}{Eqt.}{Eqts.}
\crefformat{Grammar}{Grammar #1}

\usepackage{enumitem}
\usepackage{booktabs}
\usepackage{svg}
\usepackage{listings}

\newcommand{\eg}{\textit{e.g.,} }
\newcommand{\etal}{\textit{et al.}\xspace}

\newcommand{\HF}{{Hugging Face}\xspace}

\newcommand{\TFH}{{TensorFlow Hub}\xspace}

\usepackage{bera}
\usepackage{listings}
\usepackage{xcolor}

\colorlet{punct}{red!60!black}
\definecolor{background}{HTML}{EEEEEE}
\definecolor{delim}{RGB}{20,105,176}
\colorlet{numb}{magenta!60!black}

\lstdefinelanguage{json}{
    basicstyle=\normalfont\ttfamily,
    numbers=left,
    numberstyle=\scriptsize,
    numbersep=0pt,
    showstringspaces=false,
    breaklines=true,
    frame=lines,
    backgroundcolor=\color{background},
    literate=
     *{0}{{{\color{numb}0}}}{1}
      {1}{{{\color{numb}1}}}{1}
      {2}{{{\color{numb}2}}}{1}
      {3}{{{\color{numb}3}}}{1}
      {4}{{{\color{numb}4}}}{1}
      {5}{{{\color{numb}5}}}{1}
      {6}{{{\color{numb}6}}}{1}
      {7}{{{\color{numb}7}}}{1}
      {8}{{{\color{numb}8}}}{1}
      {9}{{{\color{numb}9}}}{1}
      {:}{{{\color{punct}{:}}}}{1}
      {,}{{{\color{punct}{,}}}}{1}
      {\{}{{{\color{delim}{\{}}}}{1}
      {\}}{{{\color{delim}{\}}}}}{1}
      {[}{{{\color{delim}{[}}}}{1}
      {]}{{{\color{delim}{]}}}}{1},
}
\usepackage{scalerel}
\usepackage{tikz}
\usetikzlibrary{svg.path}

\definecolor{orcidlogocol}{HTML}{A6CE39}
\tikzset{
    orcidlogo/.pic={
        \fill[orcidlogocol] svg{M256,128c0,70.7-57.3,128-128,128C57.3,256,0,198.7,0,128C0,57.3,57.3,0,128,0C198.7,0,256,57.3,256,128z};
        \fill[white] svg{M86.3,186.2H70.9V79.1h15.4v48.4V186.2z}
        svg{M108.9,79.1h41.6c39.6,0,57,28.3,57,53.6c0,27.5-21.5,53.6-56.8,53.6h-41.8V79.1z M124.3,172.4h24.5c34.9,0,42.9-26.5,42.9-39.7c0-21.5-13.7-39.7-43.7-39.7h-23.7V172.4z}
        svg{M88.7,56.8c0,5.5-4.5,10.1-10.1,10.1c-5.6,0-10.1-4.6-10.1-10.1c0-5.6,4.5-10.1,10.1-10.1C84.2,46.7,88.7,51.3,88.7,56.8z};
    }
}

\newcommand\orcidicon[1]{\href{https://orcid.org/#1}{\mbox{\scalerel*{
                \begin{tikzpicture}[yscale=-1,transform shape]
                \pic{orcidlogo};
                \end{tikzpicture}
            }{|}}}}

\usepackage{hyperref}

\begin{document}

\newcommand{\numberOfModelHub}{5\xspace}

\newcommand{\TotalNumberOfPackages}{{15,913}\xspace}

\newcommand{\HFNumberOfPackages}{{12,401}\xspace}
\newcommand{\HFNumberOfPackagesMetadata}{{124,427}\xspace}
\newcommand{\MZNumberOfPackages}{3,245\xspace}
\newcommand{\PHNumberOfPackages}{{49}\xspace}
\newcommand{\MHNumberOfPackages}{{33}\xspace}
\newcommand{\ONNXNumberOfPackages}{{185}\xspace}

\newcommand{\TotalDataSize}{\textasciitilde{61TB}\xspace}
\newcommand{\HFDataSize}{{61TB}\xspace}
\newcommand{\MZDataSize}{{115GB}\xspace}
\newcommand{\PHDataSize}{{1.5GB}\xspace}
\newcommand{\MHDataSize}{{721MB}\xspace}
\newcommand{\ONNXDataSize}{{441MB}\xspace}


\newcommand{\PTMDatasetNPackages}{63,182\xspace}
\newcommand{\PTMDatasetPercentage}{{99.7\%}\xspace}
\newcommand{\PTMDatasetFailedPackages}{{186}\xspace}
\newcommand{\PTMDatasetFailedPercentage}{{0.3\%}\xspace}

\newcommand{\PTMDatasetNReposWithSignedCommits}{{132}\xspace}
\newcommand{\PTMDatasetPercentOfSignedCommits}{{0.208\%}\xspace}

\newcommand{\PercentOfVerifiedOrgs}{{3.188\%}\xspace}
\newcommand{\NOrganizations}{{6,243}\xspace}
\newcommand{\NVerifedOrgs}{{199}\xspace}

\newcommand{\NOfRepositoriesWithMalware}{{1}\xspace}
\newcommand{\PercentageOfRepositoriesWithMalware}{{0.002\%}\xspace}
\newcommand{\TotalRepositoriesForMalwareScanning}{{63,366}\xspace}


\newcommand{\MyTitle}[1]{}

\renewcommand{\MyTitle}{PTMTorrent: A Dataset of Package Snapshots in PTM Registries }
\renewcommand{\MyTitle}{PTMTorrent: An Open-source Dataset for Mining Pre-trained Model Packages}
\renewcommand{\MyTitle}{PTMTorrent: A Dataset of Pre-trained Machine Learning Models From Five Registries}
\renewcommand{\MyTitle}{PTMTorrent: 124,000 Pre-trained Machine Learning Models From Five Registries}
\renewcommand{\MyTitle}{PTMTorrent: A Dataset for Mining Open-source Pre-trained Model Packages}

\newcommand{\orcid}[1]{\href{https://orcid.org/#1}{\textcolor[HTML]{A6CE39}{\aiOrcid}}}

\title{\MyTitle}

        \author{
    \IEEEauthorblockN{Wenxin Jiang \IEEEauthorrefmark{1}\textsuperscript{\textsection}$^{\textsuperscript{\orcidicon{0000-0003-2608-8576}}}$\, 
    Nicholas Synovic\IEEEauthorrefmark{2}\textsuperscript{\textsection}$^{\textsuperscript{\orcidicon{0000-0003-0413-4594}}}$\, 
    Purvish Jajal\IEEEauthorrefmark{1}$^{\textsuperscript{\orcidicon{0000-0002-1199-6363}}}$\, 
    Taylor R. Schorlemmer\IEEEauthorrefmark{1}$^{\textsuperscript{\orcidicon{0000-0003-2181-5527}}}$\, 
    Arav Tewari\IEEEauthorrefmark{1}$^{\textsuperscript{\orcidicon{0000-0002-1512-858X}}}$\,\\ 
    Bhavesh Pareek\IEEEauthorrefmark{1}$^{\textsuperscript{\orcidicon{0000-0002-6885-9810}}}$\, 
    George K. Thiruvathukal\IEEEauthorrefmark{2}$^{\textsuperscript{\orcidicon{0000-0002-0452-5571}}}$\, 
    James C. Davis\IEEEauthorrefmark{1}$^{\textsuperscript{\orcidicon{0000-0003-2495-686X}}}$\
    }
    \IEEEauthorblockA{\IEEEauthorrefmark{1}Purdue University and \IEEEauthorrefmark{2}Loyola University Chicago}
}
    
\maketitle
\begingroup\renewcommand\thefootnote{\textsection}
\footnotetext{Authors contributed equally.}
\endgroup

\begin{abstract} \label{sec: abstract}
Due to the cost of developing and training deep learning models from scratch, machine learning engineers have begun to reuse pre-trained models (PTMs) and fine-tune them for downstream tasks.
PTM registries known as ``model hubs'' support engineers in distributing and reusing deep learning models.
PTM packages include pre-trained weights, documentation, model architectures, datasets, and metadata.
Mining the information in PTM packages will enable the discovery of engineering phenomena and tools to support software engineers.
However, accessing this information is difficult --- there are many PTM registries, and both the registries and the individual packages may have rate limiting for accessing the data.

We present an open-source dataset, PTMTorrent, to facilitate the evaluation and understanding of PTM packages.
This paper describes the creation, structure, usage, and limitations of the dataset.
The dataset includes a snapshot of \numberOfModelHub model hubs and a total of \TotalNumberOfPackages PTM packages.
These packages are represented in a uniform data schema for cross-hub mining.
We describe prior uses of this data and suggest research opportunities for mining using our dataset.

The \textit{PTMTorrent} dataset (v1) is available at: \url{https://app.globus.org/file-manager?origin_id=55e17a6e-9d8f-11ed-a2a2-8383522b48d9&origin_path=\%2F\%7E\%2F}.

Our dataset generation tools are available on GitHub: 
\url{https://doi.org/10.5281/zenodo.7570357}




\end{abstract}

\begin{IEEEkeywords}
Open-Source Software, 
Data Mining,
Machine learning,
Empirical software engineering
\end{IEEEkeywords}



\section{Introduction} \label{sec:Intro}
Modern software systems reuse Deep Neural Networks (DNNs) to build intelligent and adaptive systems~\cite{Amershi2019SE4MLCaseStudy, Shafiq2021LitReviewofMLinSWDevLifeCycle}.
Engineering a DNN from scratch is challenging for many reasons,
  including the variation in deep learning libraries~\cite{Pham2020AnalysisofVarianceinDLSWSystems,banna2021experience}
  and
  the high expense of training models~\cite{patterson2021carbon}.
Organizations and developers can address some of these challenges and reduce the cost and effort associated with DNN development by reusing \emph{pre-trained DNN models} (PTMs)~\cite{Tan2018DeepTransferLearningSurvey, Pan2010TransferLearning}. 
PTMs are shared via \emph{deep learning model registries}, which are modeled on traditional software package registries such as NPM~\cite{npm}. These PTM packages include reusable components, such as model architectures, weights, licenses, and other metadata.
Deep learning model registries enable engineers to develop their models with re-usability in mind~\cite{TensorFlowHubIntroduction,HuggingFacePaper2020}.
Although PTM reuse is still in its early stages, the most popular PTMs are downloaded millions of times each month~\cite{2022JiangEmpirical, Jiang2022PTMReuse}.

As PTM reuse becomes more widespread, the engineering community will benefit from research into PTM reuse practices, challenges, and tools~\cite{Jiang2022PTMReuse, 2022JiangEmpirical}.
By analogy to traditional software, mining PTM software repositories can help us understand development trends~\cite{Ray2014StudyofProgrammingLanguagesandCodeQualityinGitHub, Zampetti2017HowOSProjectsUseStaticCodeAnalysisinCIPipelines, Gousios2014PullBasedSWDevelopment} and usage patterns~\cite{anbalagan2009predicting, Malinen2015UserParticipationinOnlineCommunities}. 
However, mining the software repositories associated with PTM packages is difficult for three reasons related to \emph{data availability}.
First, researchers must look in many places --- PTM packages are distributed across many competing PTM registries~\cite{2022JiangEmpirical}.
Second, researchers must access the packages --- PTMs include complex DNN models and weights with sizes over 1 TB, and access to these packages may be hindered by throttling or rate limiting~\cite{HFDoc}.
Third, for scientific replicability, this large-scale data needs to be hosted long-term.

To enable mining of PTM packages, we share \emph{PTMTorrent}, the first many-hub dataset of PTM packages.
PTMTorrent contains \TotalNumberOfPackages PTMs from \numberOfModelHub different PTM registries identified in our prior work~\cite{2022JiangEmpirical}: \HF~\cite{HuggingFaceWeb1}, Model Zoo~\cite{ModelZooWeb},
PyTorch Hub~\cite{PytorchHub}, ONNX Model Zoo~\cite{ONNXModelZoo}, and Modelhub~\cite{ModelhubWeb}.
Our dataset is hosted on a high-performance storage system (HPSS) maintained by Purdue University's Research Computing center.
The dataset includes the metadata of each PTM and the package histories for each GitHub repository.
These packages are represented in a uniform data schema for cross-hub mining.
Out dataset supports many directions for further research, including studies of the PTM supply chain, PTM package evolution, PTM mining tools, and DNN architectural trends.


{
\small
\renewcommand{\arraystretch}{0.9}
\begin{figure*}[t]
    \centering
    \includegraphics[width=0.7\textwidth]{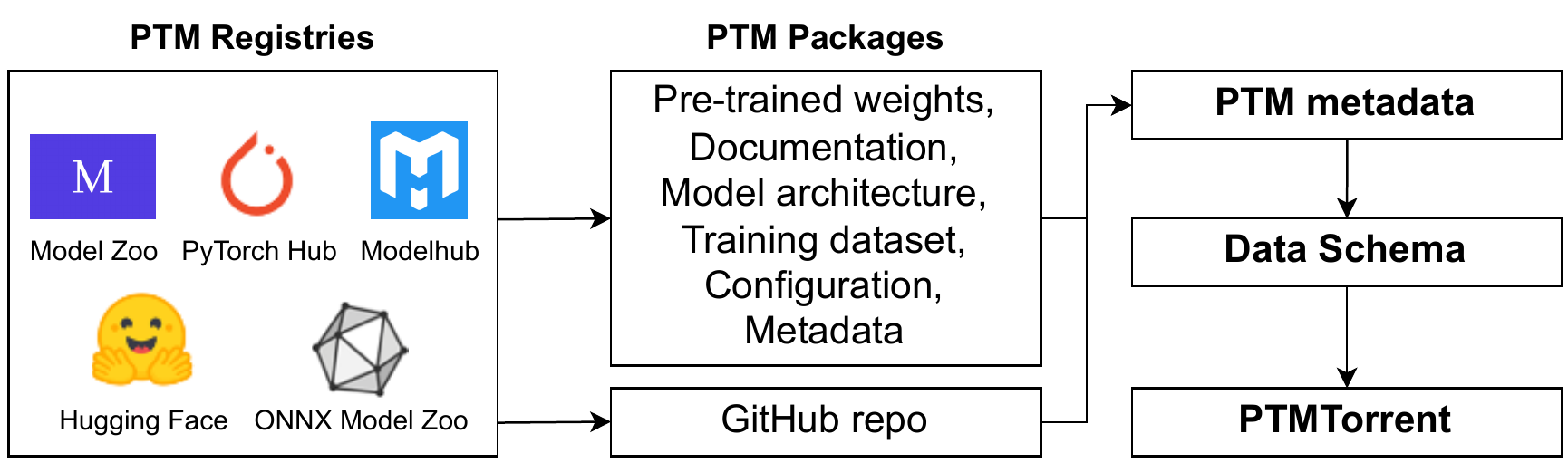}
    \caption{
    \small
    Data collection and preprocessing workflow for PTMTorrent. We standardize the PTM metadata by using a data schema, collecting it from PTM packages and the corresponding GitHub repository.
    }
    \label{fig:DataCollection}
\end{figure*}
}

\section{The PTMTorrent Dataset} \label{sec:Dataset}

\subsection{Data Source} \label{sec:PTMTorrent-DataSource}

In prior work we mapped the major model hubs and indicated that there exist open, gated, and commercial hubs~\cite{2022JiangEmpirical}.
Open and gated hubs tend to be larger and more widely used because they accept contributions from anyone, and can be accessed by anyone.
Commercial hubs are offered by individual companies to share vetted models with their clients.
Due to the limited access to commercial model hubs, we only provide a snapshot of the open hubs (\HF) and some of the gated hubs (Model Zoo, PyTorch Hub, Modelhub, and ONNX Model Zoo).

The PTMTorrent dataset contains the repository histories of \TotalNumberOfPackages PTM packages available as of January 2023.
They are provided as complete git clones, resulting in a compressed footprint of \TotalDataSize. 
Each PTM package was cloned at its most recent version, including the model card, architecture, weights, and other information provided by the maintainers (\eg training configuration, hyper-parameters).

\cref{fig:DataCollection} indicates the collection and preprocessing approaches of our dataset.

We collected PTM packages from all open and gated model hubs per Jiang \etal~\cite{2022JiangEmpirical}, excluding \TFH because it does not support version control features.
We downloaded all PTM packages from Model Zoo, PyTorch Hub, ONNX Model Zoo, and Modelhub.
Due to the size of Hugging Face, we downloaded only the top 10\% most-downloaded PTMs.\footnote{Although we collected a small amount of the full Hugging Face registry, this ``top 10\%'' snapshot includes all Hugging Face PTMs with over 30 downloads.}
Overall, our dataset contains \TotalNumberOfPackages packages from \numberOfModelHub PTM registries, distributed as described in~\cref{tab:DatasetContents}.

{
\small
\renewcommand{\arraystretch}{0.5}
\begin{table}[h]
\centering
\caption{
    Details about the PTMTorrent content for each of the \numberOfModelHub model registries we collected.
    }
\begin{tabular}{lcccc}
\toprule
          \textbf{Name} & \textbf{\# Models} & \textbf{Data Size} \\
\midrule
  Hugging Face~\cite{HuggingFaceWeb} &    \HFNumberOfPackages  &  \HFDataSize\\
\\
Model Zoo~\cite{ModelZooWeb}  &         \MZNumberOfPackages & \MZDataSize\\
\\
PyTorch Hub~\cite{PytorchHub} &         \PHNumberOfPackages &  \PHDataSize\\
\\
ONNX Model Zoo~\cite{ONNXModelZoo}  &         \ONNXNumberOfPackages &   \ONNXDataSize\\
\\
Modelhub~\cite{ModelhubWeb}  &        \MHNumberOfPackages &   \MHDataSize\\
\midrule
\textbf{PTMTorrent}  &       \textbf{\TotalNumberOfPackages} & 
\textbf{\TotalDataSize}\\
     
\bottomrule
\end{tabular}
\label{tab:DatasetContents}
\end{table}
}

\subsection{Data Schema}
\cref{fig:DataSchema} shows the overview of the data schema we used to standardize the dataset. 
We extracted common entities into a general PTM schema.
Each PTM registry has some custom features, so we customized the schema slightly for each model registry.
The full data schema is encoded following the JSON Schema format,\footnote{See \url{https://json-schema.org/draft/2020-12/json-schema-core.html}} and is available in the GitHub repository associated with this project.





\subsection{Data Storage}
As shown in \cref{tab:DatasetContents}, the entire PTMTorrent dataset (v1) needs \TotalDataSize of storage space. 
A cost-effective storage system is required to serve this dataset.
Commercial services are cost-prohibitive at this scale, \eg we estimated a monthly cost of over \$1000 to store and serve this dataset from Amazon Web Services.
We opted instead for an internal resource available at Purdue University: the Purdue Fortress tape-based hierarchical storage system.\footnote{For more information about Fortress, see \url{https://www.rcac.purdue.edu/knowledge/fortress/overview}. Our GitHub repository includes a guide on how to access data stored in Globus.} 
To facilitate external distribution of our dataset, we offer a Globus share~\cite{chard2015globus} named \textit{PTMTorrent}.

\subsection{Maintainability and Extensibility}

The sizes of PTM registries are increasing rapidly. For example, \HF provided \PTMDatasetNPackages public PTM packages on August 2022, and now it provides \HFNumberOfPackagesMetadata packages.
We believe the number of open-source PTM packages will increase in the foreseeable future. 
Therefore, maintainability and extensibility are two important properties of PTMTorrent.

The PTMTorrent dataset is designed to be maintainable by re-running our scripts to gather any additional changes that may have been made to the PTM registry since its last collection. 
Expect a biannual update.

For extensibility, new model hubs can be incorporated into the dataset.
We follow an open-source model and will review Issue and Pull Request contributions on GitHub.
The PTMTorrent data schema captures most elements of a PTM package, though some specialization is needed.
The downloaders for a new model hub can be developed based on the examples of the already-supported model hubs in our open-source data collection tools.
An extender must provide 2-4 scripts following the pattern we used on the other hubs.

{
\begin{figure*}[t]
    \centering
    \includegraphics[width=0.75\textwidth]{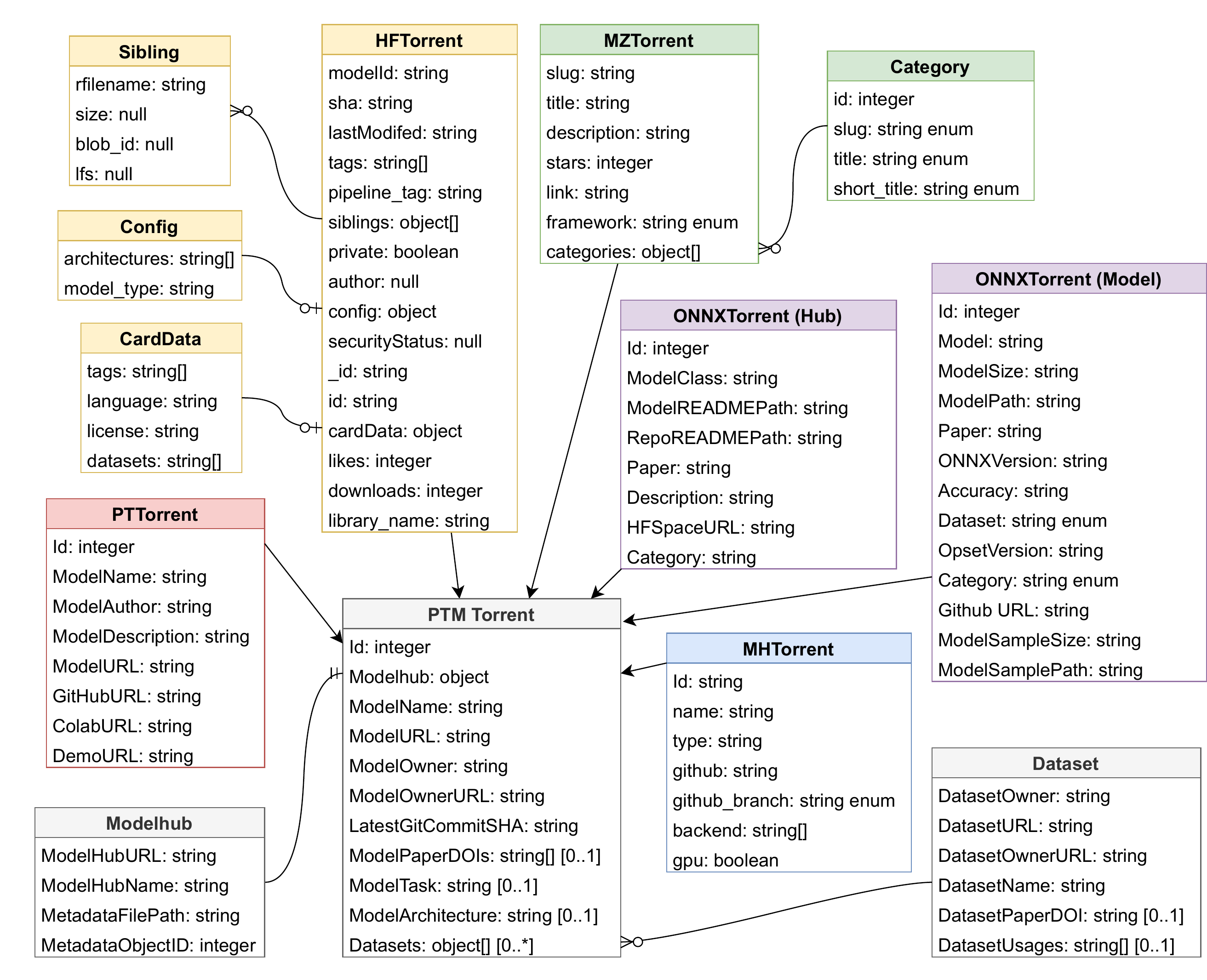}
    \caption{
    \small
    An overview of PTMTorrent's data schema. Each model hub shares a general schema (\textit{grey boxes}), with hub-specific data stored in customized schema (\textit{colored boxes}). The full schema is available in JSON in the dataset generation repository. 
    }
    \label{fig:DataSchema}
\end{figure*}
}

\section{Originality and Relevance}
\label{sec:Originality}

Prior works have extracted information from open-source projects to a dataset and provide it for future analysis, such as GHTorrent~\cite{Gousios2012GHTorrent}, TravisTorrent~\cite{Beller2017TravisTorrent}, and RTPTorrent~\cite{Mattis2020RTPTorrent}.
These datasets can be used for further mining software repository researches and help the community better understand open-source software projects~\cite{Beller2017ExplorativeStudyofTravisCI, Zampetti2017HowOSProjectsUseStaticCodeAnalysisinCIPipelines, Gousios2014PullBasedSWDevelopment, Elsner2021RegressionTestOptimizationinCI}.

Similarly, our dataset captures the open-source PTM packages from many model hubs.
The structure of our dataset imitates prior datasets that were focused on traditional open-source software~\cite{Gousios2012GHTorrent, Mattis2020RTPTorrent}.
Compared to prior work, PTMTorrent focuses on PTM packages, including the metadata, architecture, dataset, and performance metrics.
Our dataset provides a way for users to efficiently download and access large amount of data on PTM packages and relevant repositories. 



\section{Usage Examples}

\subsection{Prior Usage in the Literature}
In prior work, we used a part of PTMTorrent (the Hugging Face part) to measure potential risks in the \HF model registry~\cite{Jiang2022PTMReuse}. 
We measured the dependencies of model architecture and datasets, PTM documentation, and GPG commit signing in \HF PTMs.
Our analysis identified potential software supply chain concerns facing PTM reusers, including spoofing, tampering, and repudiation. 

In prior work, we also used metadata from \HF to measure model discrepancies and maintainers' reach~\cite{2022JiangEmpirical}.
Our analysis showed that existing defenses appear insufficient for ensuring the security of PTMs.

The PTMTorrent dataset provides more opportunities for mining PTM data by covering more PTM registries and providing greater structure.
We believe that these large amount of PTM packages can be analyzed in similar ways as traditional packages~\cite{Zimmermann2019SecurityThreatsinNPMEcosystem, Zahan2022WeakLinksinNPMSupplyChain, Decan2018SecurityVulnerabilitiesinNPMDependencyNetwork}.


\subsection{Applying an Existing MSR Tool}

Since PTMTorrent consists of git repositories, it is possible to use existing software repository mining tools on the PTM packages. 
Our GitHub repository includes a demonstration of this.
We used our PRIME tool~\cite{synovic_snapshot_2022} to analyze software process metrics on a subset of the dataset.

\section{Limitations} \label{sec:Limitations}

PTMTorrent is incomplete.
It is biased towards the top 10\% most-downloaded PTMs in HuggingFace (though this is almost all PTMs with any downloads, cf.~\cref{sec:PTMTorrent-DataSource}).
There are other model hubs, such as Papers With Code~\cite{PapersWithCode}, PINTO Model Zoo~\cite{PINTOModelZoo}, and Jetson Zoo~\cite{JetsonZoo}. 
Beyond these, there are other deep learning-specific registries that lack versioning or packaging features.
The initial PTMTorrent release provides PTMs from model hubs that are similar to traditional software packages, as defined by Jiang \etal~\cite{2022JiangEmpirical}.
We leave their capture for future work.


Another limitation of our data is the non-standardized granularity.
The current version of PTMTorrent lacks detailed metadata and does not provide uniform information, \eg datasets, model architectures. During the data collection, we notice that the provided information from PTM registries can be quite different and we use customized data schemas for each PTM registry. As a result, it is difficult to analyze all the PTM packages under the same umbrella when using our dataset.

For example, \HF provides detailed documentation and structured metadata, as well as relevant configuration files for each PTM, while ONNX Model Zoo provides PTM metadata through unstructured Markdown files. Thereby making metadata extraction challenging. 
To mitigate this problem, we have a parent data schema for all the PTM registries and child schemas for each specific registry that represents their custom data.

\section{Future Work} \label{sec:Future Work}
In addition to the risk measurements presented by Jiang \etal~\cite{Jiang2022PTMReuse,2022JiangEmpirical}, the PTMTorrent dataset can be used in different ways.
We suggest three research directions:
  PTM supply chain analysis,
  tools for PTM reuse,
  and
  mining tool development.

\subsection{Supporting Future PTM Supply Chain Analysis}
Prior work has focused on understanding the characteristics of package registries and their supply chains.
Zimmermann \etal analyzed the metadata of NPM packages and identified the potential threats on downstream users~\cite{Zimmermann2019SecurityThreatsinNPMEcosystem}.
Ladisa \etal proposed an attack taxonomy on open-source supply chains, including code contributions to package distribution~\cite{Ladisa2022TaxonomyofAttackonOSSSupplyChains}.
Similar studies are also important in PTM supply chain alongside studies focused on PTM-specific aspects.
We propose that future studies can analyze PTMTorrent dataset to understand the characteristics of the PTM supply chain, including the dependency analysis~\cite{Zimmermann2019SecurityThreatsinNPMEcosystem}, vulnerabilities~\cite{Alfadel2021SecurityAnalysisinPythonPackages}, and code knowledge transfer~\cite{Ma2021OpenSourceVCSData}.
Recent advances in AI, such as ChatGPT~\cite{ChatGPT}, that clearly build upon composing various PTMs strongly suggest that being able to study how PTMs and are composed to build more complex systems (a trait shared with traditional software) will become more important.
We hope our dataset will aid in performing such analyses.

\subsection{Expanding PTM Model Registry Analysis}
Researchers can extract more information from these model registries by reusing or developing software metrics for PTM packages, including provenance, reproducibility, and portability~\cite{Jiang2022PTMReuse}. PTM registries can help us develop comprehensive attributes and provide these details in the PTM dashboard, similar to the measured attributes from NPM~\cite{npmsAttribute} and PyPi~\cite{pypi}. 

Our prior study has indicated that engineers can have trouble finding the best PTM that matches their requirements, and it can therefore be hard to identify the portability and reproducibility of the open-source PTMs~\cite{Jiang2022PTMReuse}. 
Montes \etal shows that there exist notable discrepancies among different model zoos~\cite{montes_discrepancies_2022}. 
With more detailed and comprehensive metadata provided for each PTM and the corresponding usage patterns on downstream tasks, it will be possible to develop a recommender
system to help engineers find the right set of PTMs for a given application and requirements~\cite{Robillard2010RecommendationSystems4SE}. PTM Registry contributors can develop sophisticated visualization tools---with the aid of our dataset---that help PTM users understand the strengths and limitations of each model.

\subsection{Furthering the State of Mining Tool Development}
Given the lack of standardization among different PTM registries (\cref{sec:Limitations}), it was challenging to standardize all the metadata.
PTMTorrent may not have everything needed for every type of analysis.
Researchers can 
augment the dataset during the data collection and processing stage for other subsequent mining needs. 
We have included the relevant GitHub pages of each PTM in out dataset, and therefore the extraction can be done either based on the provided documentation from PTM registries~\cite{Slankas2013AutomatedExtractionofNonfunctionalRequirements} or source code from the underlying repositories~\cite{Allamanis2016AttentionNetwork4ExtremeSummarizationofSourceCode}.

\ifANONYMOUS
\else
\section{Acknowledgements}
This work was supported by gifts from Google and Cisco and by NSF awards \#2107230, \#2229703, \#2107020, and \#2104319.
\fi

\raggedbottom
\pagebreak
\balance

\bibliographystyle{refs/IEEEtran}
\bibliography{main}
%
\raggedbottom
\pagebreak
\balance

\end{document}